\documentclass[11pt,twoside]{article}

\usepackage{asp2006}
\usepackage{epsf}
\usepackage{psfig}
\usepackage{lscape}

\def\gs{{_>\atop^{\sim}}}
\def\ergss{ergs s$^{-1}$\/}

\begin{document}
\title{Time-Evolving Photoionization: the Thin and Compact X-Ray Wind of NGC~4051}   
\author{Fabrizio Nicastro$^1$, Martin Elvis, Nancy Brickhouse}   
\affil{Harvard-Smithsonian CfA, Cambridge, Massachusetts, USA; 
$^1$OAR-INAF, Monte Porzio Catone (RM), Italy}
\author{Yair Krongold, Luc Binette}
\affil{Instituto de Astronomia - UNAM, Mexico D.F., Mexico}
\author{Smita Mathur}
\affil{Astronomy Department of Ohio State University, Columbus, Ohio, USA}

\begin{abstract} 
We discuss the power of time-evolving photoionization as a diagnostic tool to measure 
the electron density of photoionized gas.  
We apply this technique to a XMM-{\em Newton} observation of the ionized 
absorber in the Seyfert 1 galaxy NGC~4051, and present the first measurements 
of the its volume density, its distance from the central ionizing source, and so 
its mass outflow rate. By extrapolating these measurements to high-luminosity, large 
black hole mass, quasars, we speculate that AGN winds can play important roles both in the
 AGN-host-galaxy and AGN-IGM feedback processes. 
\end{abstract}


\section{Ionized AGN Winds: Open Questions}
Ionized X-ray absorbers ({\em Warm Absorbers}: WAs) have been observed in 
the spectra of $\gs 50$ \% of Seyfert 1s (e.g. 
Crenshaw, Kraemer \& George 2003) and quasars (Piconcelli 
et al. 2005). Such high detection rates, 
combined with evidence for transverse flows (Mathur et al. 1995;
Crenshaw, Kraemer \& George 2003) suggest that WAs are actually
ubiquitous in AGN, but become directly visible in absorption only when our line 
of sight crosses the outflowing  material. 
Despite their ubiquity, very little is still know about their physical 
state, dynamical strength and geometry. 
The most fundamental  question is where do quasar winds originate?
Addressing this question requires the independent determination of a quantity which is 
not directly observable: the electron density $n_e$ of the outflowing material. 

\section{Time Evolving Photoionization} 
Recent studies have mostly shown that WAs are made up of 
just a few distinct physical components (e.g. Krongold et al., 2007 - K07 - and 
references therein). However, for each of these components, 
only average estimates of the product $(n_e R^2)$ could be derived, from time-averaged 
spectral analyses. This is due to the intrinsic degeneracy of $n_e$ and $R$ in the 
equation that defines the two observables: the ionization parameter of the 
gas $U_x = Q_x / (4 \pi R^2 c n_e)$ and the luminosity of ionizing photons $Q_x$. 

An unambiguous method to remove this degeneracy is to monitor the
response of the ionization state of the gas in the wind to changes of
the ionizing continuum (Krolik \& Kriss 1995; Nicastro et al. 1999: hereinafter N99). 
Time-evolving photoionization equations define the Photoionization Equilibrium Time scale, 
$t_{eq}$, that measures the time necessary for the gas to reach photoionization 
equilibrium with the ionizing continuum (N99).  
This time-scale depends explicitly on the
electron density $n_e$ in the cloud, during both increasing and decreasing 
ionizing continuum phases (N99). 
Non-equilibrium photoionization models can therefore be used to measure the 
density of the gas $n_e$ and, hence, its distance from the ionizing source. 

Here, we apply this technique to a high S/N XMM-{\em Newton} observation 
of the low luminosity ($L_{bol} = 2.5 \times 10^{43}$ \ergss, Ogle et al. 2004), 
low black hole mass (M$_{BH} = 1.9 \times 10^6$ M$_{\odot}$, Peterson et al. 2004), 
and rapidly ($\sim$ 1 hour) and highly (factor of $\sim 10$, McHardy et al. 2004) 
X-ray variable Narrow Line Seyfert 1 NGC~4051. 

\section{The Variable Ionized Absorber of NGC~4051}
NGC 4051 was observed for $\sim$117 ks with the XMM-Newton
Observatory, on 2001 May 16-17. The source varied by a factor of a few on 
timescales as short as 1~ks, and  by a factor of $\sim$12 from minimum
to maximum flux over the whole observation (Fig. 1a).
Details of the data reduction and analysis can be found in K07. 

Two physically distinct but dynamically coincident ($v_{out} = 500$ km s$^{-1}$) 
WA phases were found in the average RGS spectrum 
of NGC~4051: a high- and a low-ionization phase (HIP and LIP). 

\begin{figure}[!ht]
\plottwo{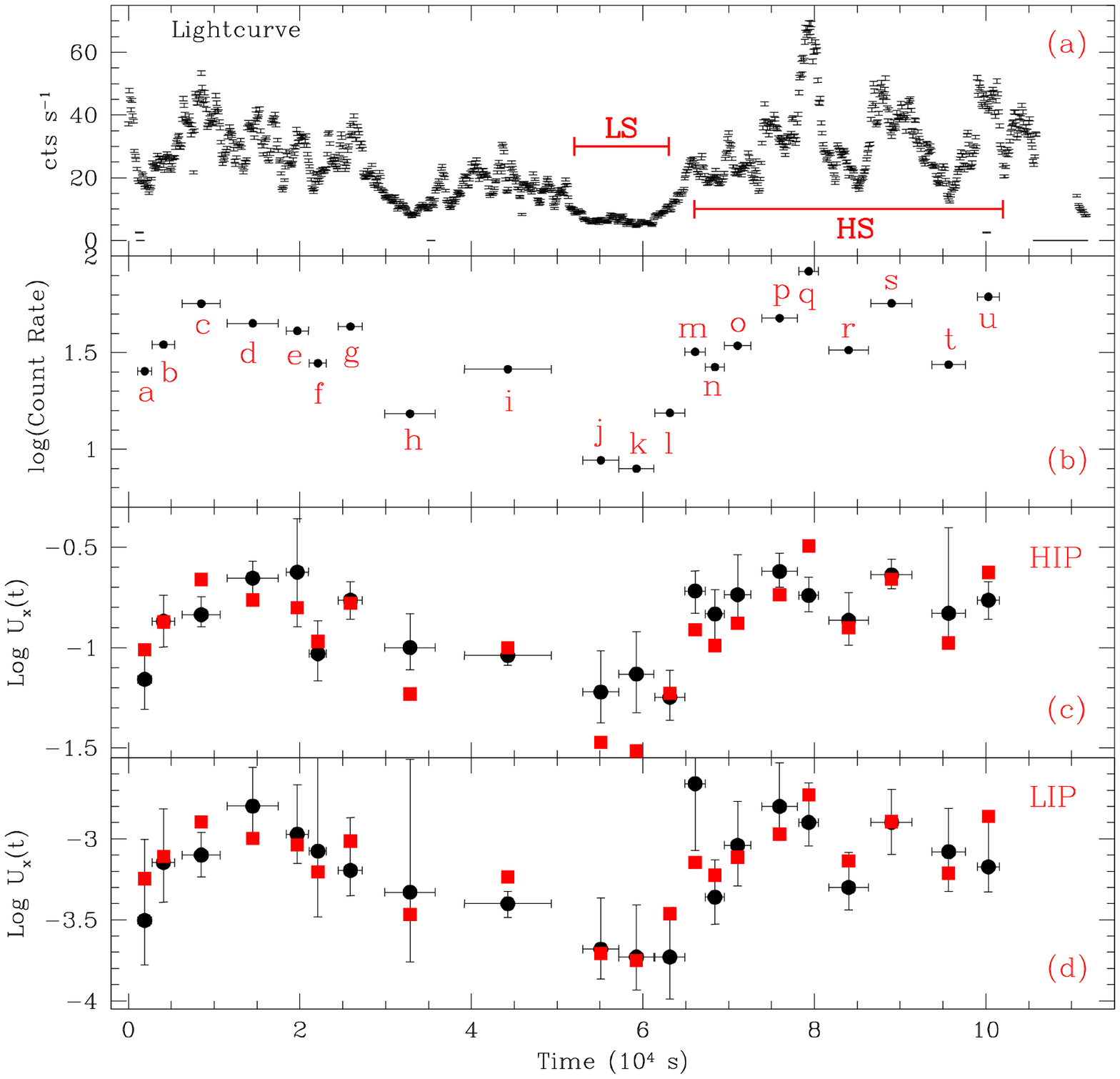}{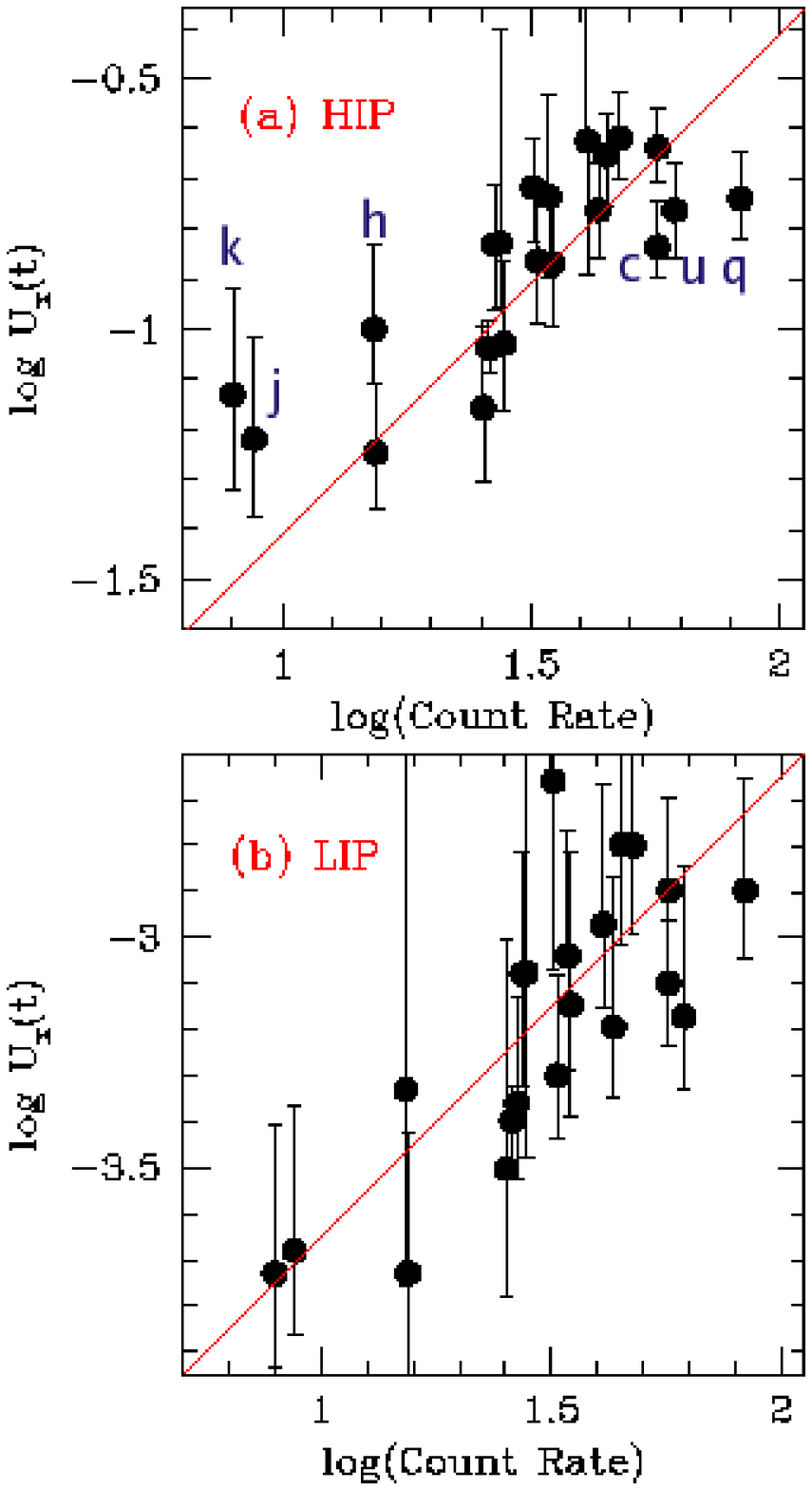}
\caption{(left) Lightcurve of NGC 4051 in bins of 100 s [panel (a)];  
Log of the count rate vs. time, for the 21 ``flux states'' used in this analysis 
[panel (b)]; Log of the ionization parameter of HIP [panel (c)] and LIP [panel (d)] 
as a function of time: filled squares are the expected value of $U_X$ at  equilibrium. 
Figure 2. \hspace{0.2cm} (right) Log$U_X(t)$ vs. log of the source count rate 
for the HIP [panel (a)] and the LIP [panel (b)].} 
\end{figure}

\subsection{Following the Time Variability of HIP and LIP}
To study the response of the WAs of NGC~4051 to ionizing flux changes, 
we performed a time-resolved spectral analysis of 21 
EPIC-PN spectra extracted from 21 distinct continuum levels ({\em a - u},
see Fig. 1b)
\footnote{We checked, on average Low- and High-State (LS and HS in Fig. 1a) 
RGS and EPIC-PN spectra, that both detectors allowed the detection of WA  
opacity variations, and gave consistent results (see K07)}
. For both absorbing components the derived $U_X$ values  follow closely
the source continuum lightcurve [compare panel (c) and (d) with panel (b) of Fig. 1], 
clearly indicating that the gas is responding quickly to the changes in the 
ionizing continuum. 

To study more quantitatively how these changes are related to the changes in the continuum, 
we show in Fig. 2a,b the log of 
the source count rate (log$C(t)$) vs. log$U_X(t)$, for the HIP
[panel (a)] and the LIP [panel (b)]. 
For most of the points of the HIP and for all the points of the LIP
(within 2$\sigma$), log$C(t)$ correlates with log$U_X(t)$ tightly, which allows us to 
derive robust estimates of the quantity ($n_e R^2$) for both components 
(best fit lines in Fig. 2a,b, Table 1). 

Fig. 2a,b shows that the LIP is always in photoionization equilibrium 
with the ionizing flux, while the HIP deviates from photoionization equilibrium 
during periods of extreme flux states. 
Unlike the LIP, then (for which only an upper limit on $t_{eq}$, 
and so a lower limit on $n_e$, can be estimated), the behavior of the HIP allows us 
to set both lower and upper limits on $t_{eq}(HIP)$, and so on $n_e(HIP)$. 

\setcounter{figure}{2}
\begin{figure}
\plotone{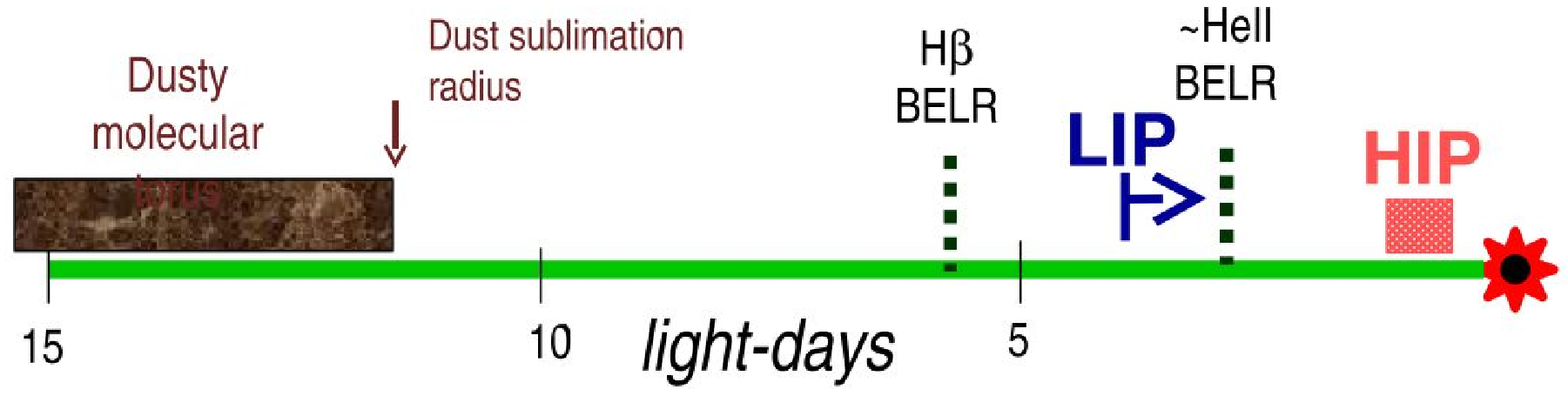}
\caption{Location of features in the nuclear environment of NGC~4051 on a light-day scale.}
\end{figure}

\begin{table}
\caption{Physical and dynamical parameters of HIP and LIP of NGC~4051}
\footnotesize
\begin{tabular}{|lccccccc|}
\hline
Abs. & $(n_eR^2)$ & $n_e$ & P$_e$ & $R$ & $\Delta R$ & $(\Delta R / R)$ & $\dot{M}_{out}
/\dot{M}_{in}$ \\
& $10^{38}$ cm$^{-1}$ & $10^{7}$ cm$^{-3}$ & $10^{12}$ K cm$^{-3}$ & $10^{15}$ cm & 
$10^{14}$ cm & & \% \\
\hline
HIP & $(0.38 \pm 0.05)$ & (0.58-2.1) & (2.9-10.5) & (1.3-2.6) & (1.9-7.2) &  (0.1-0.2) & 
(2-3)\\
LIP & $(66 \pm 3)$  & $>8.1$ & $> 2.4$ & $< 8.9$ & $< 0.09$ & $< 10^{-3}$ & $< 2$ \\
\hline
\end{tabular}
\end{table}

\section{Results}
From these independent estimates of $n_e$ and $(n_e R^2)$, 
we can derive limits on $R$ and so evaluate all physical and dynamical parameters
of the gas, including the mass outflow rate. 
Table 1 summarizes our findings. Fig. 3 shows a schematic view of the location of the 
ionized outflow in NGC~4051. We find that: 

\noindent 
(1) time-evolving photoionization models are key diagnostics to measure the electron 
density in gas photoionized by variable X-ray sources; 

\noindent 
(2) the two X-ray WAs of NGC~4051 are both close to 
photoionization equilibrium, dense, compact and possibly in pressure balance 
(Fig. 3, Table 1); 

\noindent 
(3) the Narrow Emission Line Region (NELR), molecular torus and any radial continuous 
flow, are all ruled out as origin of the WAs of NGC~4051 (Fig. 3); 

\noindent 
(4) a static spherical configuration for the WAs of NGC~4051 is ruled out, 
and a dynamical but stationary spherical configuration (i.e. expulsion of 
thin spherical shells at regular time intervals) is highly unlikely, 
since it requires extremely fine tuning not to degenerate in a radial continuous flow; 

\noindent 
(5) the next simplest geometrical configuration, for the WAs of NGC~4051, 
is that of geometrically thin cone sections, originating in the accretion disk at a 
distance consistent with that of the high-ionization BLRs (Fig. 3; Elvis, 2000); 

\noindent 
(6) if this configuration  applies to all AGN, then the presence or absence of
WAs in the X-ray spectra of type 1 AGNs is explained simply by orientation,
with WAs being visible only in those objects seen between edge-on and an angle
equal to the disk-outflow angle;

\noindent 
(7) the by-conical configuration is consistent with all our findings 
and provides reasonable values for the mass outflow rates of the WA of NGC~4051, 
relative to the source accretion rate: $\dot{M}_{out} = 2-5 $\% $\dot{M}_{in}$ (Table 1); 

\noindent 
(8) these values, if extrapolated to high luminosity QSOs, suggest that 
WAs may play an important role in locally feeding the IGM with enriched material.  

Finally, we note that a radial extension of disk winds toward larger disk radii, 
could produce cold absorption, as seen in type 2 AGNs (Risaliti, Elvis \& Nicastro, 
2002): in such objects WAs would be masked by the cold absorption, but the ionized wind 
would still shine in X-rays and could produce the photoionized X-ray cones observed
in several Seyfert 2s (Guainazzi, this conference).

\acknowledgements F.N. acknowledges support by NASA grant NNG05GK47G.

\end{document}